\documentclass[fleqn,twoside]{article}
\usepackage{espcrc2}

\usepackage{graphicx}
\usepackage{epsfig}
\usepackage{amssymb}
\usepackage[figuresright]{rotating}

\newcommand{\AmS}{{\protect\the\textfont2
  A\kern-.1667em\lower.5ex\hbox{M}\kern-.125emS}}

\newcommand{\beq}{\begin{equation}}
\newcommand{\eeq}{\end{equation}}
\newcommand{\beqn}{\begin{eqnarray}}
\newcommand{\eeqn}{\end{eqnarray}}
\newcommand{\bea}[1]{\beq\begin{array}{#1}}
\newcommand{\eea}{\end{array}\eeq}

\title{\vspace{-2.5cm}
       {\normalsize DESY 02--135}    \\[-0.2cm]
       {\normalsize ITEP-LAT/2002--16}   \\[-0.2cm]
       {\normalsize KANAZAWA 02--26}   \\[0.8cm]
Thermodynamics and heavy quark
potential in $N_f=2$ dynamical QCD \thanks{Talks given by V.~Bornyakov and Y.~Nakamura at Lattice 2002, MIT, Cambridge MA, USA.}}

\author{V. Bornyakov \address[KU]{Institute for Theoretical Physics, Kanazawa University,
Kanazawa 920-1192, Japan}$^,$\address[ITEP]{Institute for Theoretical and Experimental
Physics, B.Cheremushkinskaya 25, Moscow 117259, Russia},
Y. Nakamura \addressmark[KU],
      M. Chernodub \addressmark[KU]$^,$\addressmark[ITEP],
Y. Koma \addressmark[KU], Y. Mori \addressmark[KU],
M. Polikarpov \addressmark[ITEP], \\
G. Schierholz \address{NIC/DESY Zeuthen, Platanenallee 6, 15738 Zeuthen,
Germany and \\ ~~Deutsches Elektronen-Synchrotron DESY, D-22603 Hamburg, Germany},
A. Slavnov \address{Steklov Mathematical Institute, Vavilova 42, 117333 Moscow, Russia},
H. St\"uben \address{Konrad-Zuse-Zentrum f\"ur Informationstechnik
Berlin, D-14195 Berlin, Germany},
T. Suzuki \addressmark[KU], P. Uvarov \addressmark[ITEP], A. Veselov \addressmark[ITEP]
}

\begin{document}

\begin{abstract}
We study $N_f=2$ lattice QCD with
nonperturbatively improved Wilson fermions at finite temperature
on $16^3 \cdot 8$ lattices.
We determine the transition temperature at $\frac{m_{\pi}}{m_{\rho}} \sim 0.8$
and lattice spacing as small as $a \sim 0.12$fm. The string breaking at
$T < T_c$ is also studied.
We find that the static potential can be fitted by a simple expression
involving string model potential at finite temperature.
\vskip -3mm
\end{abstract}

\maketitle

\section{Introduction}
Recent  studies of $N_f=2$ lattice QCD at finite temperature with
improved actions have provided consistent estimates of $T_{c}$.
Bielefeld group employed improved staggered fermions and improved
gauge field action \cite{Karsch:2000kv}. CP-PACS collaboration
used improved Wilson fermions with mean field improved $c_{sw}$
and improved gauge field action \cite{AliKhan:2000iz}. Both groups
were able to estimate $T_c$ in the chiral limit and their values
are in a good agreement. Still there are many sources of
systematic uncertainties and new computations of $T_c$  with
different actions are useful as an additional check. We made first
large scale simulations of the nonperturbatively $O(a)$ improved
Wilson fermion action.  Moreover we performed simulations with the
lattice spacing $a$ much smaller than in studies by Bielefeld and
CP-PACS groups. Our small lattice spacing helps us to determine
parameters of the static potential in full lattice QCD at finite
temperature. Our other goal is to study the vacuum structure of
the full QCD at $T>0$, in particular a relation between string
breaking observed with the help of the Polyakov loop correlator at
$T<T_c$ and abelian monopoles.

The fermionic action we employ is of the form
\begin{equation}
S_F = S^{(0)}_F - \frac{\rm i}{2} \kappa\, g\
c_{sw} a^5
\sum_x \bar{\psi}(x)\sigma_{\mu\nu}F_{\mu\nu}\psi(x),
\label{one}
\end{equation}
where $S^{(0)}_F$ is the original Wilson action, $c_{sw}$
is determined nonperturbatively \cite{Jansen:1998mx}.
We use Wilson gauge field action.

We took advantage of availability of $T=0$ results obtained with the 
action~(\ref{one}) by UKQCD and QCDSF collaborations
\cite{Booth:2001qp} to fix the physical scale and
$\frac{m_{\pi}}{m_{\rho}}$ ratio. Studies made by these
collaborations also confirmed that $O(a)$ lattice artifacts are
suppressed as expected \cite{Allton:2002sk}. We then may hope
that lattice discretization errors of our results are small.
So far only $N_t=4$ and 6 finite temperature results obtained with
this action are available~\cite{Edwards:1999mm}. These results were obtained
at rather large quark mass ($m_{\pi}/m_{\rho}>0.85$) and lattice spacing.

\begin{table}[thb]
\begin{center}
\begin{tabular}{|c|c|c|c|}
\hline
\multicolumn{2}{|c|}{$\beta=5.2$} & \multicolumn{2}{|c|}{$\beta=5.25$} \\ \hline
 $\kappa$ & \# of traj. & $\kappa$ & \# of traj. \\ \hline
 0.1330   & 3410       & 0.1330   & 1540   \\
 0.1335   & 1350       & 0.1335   & 1600  \\
 0.1340   & 2100       & 0.13375  & 9225   \\
 0.1343   & 2562       & 0.1339   & 12470  \\
 0.1344   & 3631       & 0.1340   & 10200  \\
 0.1345   & 3507       & 0.1341   & 2608   \\
 0.1346   &  350       & 0.13425  & 5155   \\
 0.1348   & 1515       & 0.1345   & 2650   \\
 0.1355   & 1801       & 0.1350   & 1780   \\
 0.1360   & 3699       &          &        \\
 \hline
\end{tabular}
\end{center}
\caption{\vskip -24pt \hskip 11.5mm . Simulation statistics.}
\label{table}
\vskip -6mm
\end{table}

Since $c_{sw}$ is only known nonperturbatively for $\beta \geq 5.2$,
we have to work at rather large
values of $\beta$. Thus to study transition at smaller values
of the quark mass we need to go to larger
$N_t$ values. This dictates our choice $N_t=8$. We choose spatial extension
of the lattice $N_s=16$ as a compromise between computational burden
and need to reduce finite size effects.

It is known that in quenched QCD the order parameter of the finite
temperature phase transition is Polyakov loop and corresponding
symmetry is global $Z(3)$ symmetry. 
In chiral QCD the order
parameter of the chiral symmetry breaking transition is the chiral
condensate $\langle \bar{\psi}\psi\rangle $. There is no phase
transition at intermediate values of the quark mass, only
crossover.
As numerical results show \cite{Karsch:2000kv} both
order parameters can be used to locate the transition point. We
use only Polyakov loop leaving computation of the chiral
condensate for future studies.

\section{Simulation details}
We use Hybrid Monte Carlo algorithm with parameters $\delta\tau=0.0125$,
$ n_\tau=20$,
with acceptance rate about 70\%. Simulations were done on Hitachi
SR8000 at KEK and MVS 1000M at RAS, Moscow. We needed from 1000
($\tau=250$) to 3000 ($\tau=750$) trajectories for thermalization,
depending on $\kappa$ and $\beta$. Our simulations were performed
for $\beta=5.2,\, 5.25$. The values of $\kappa$ and  corresponding
number of trajectories are listed in Table 1.

\begin{figure}[thb]
\hbox{
\epsfxsize=6.cm
\hspace{0.cm} \epsfbox{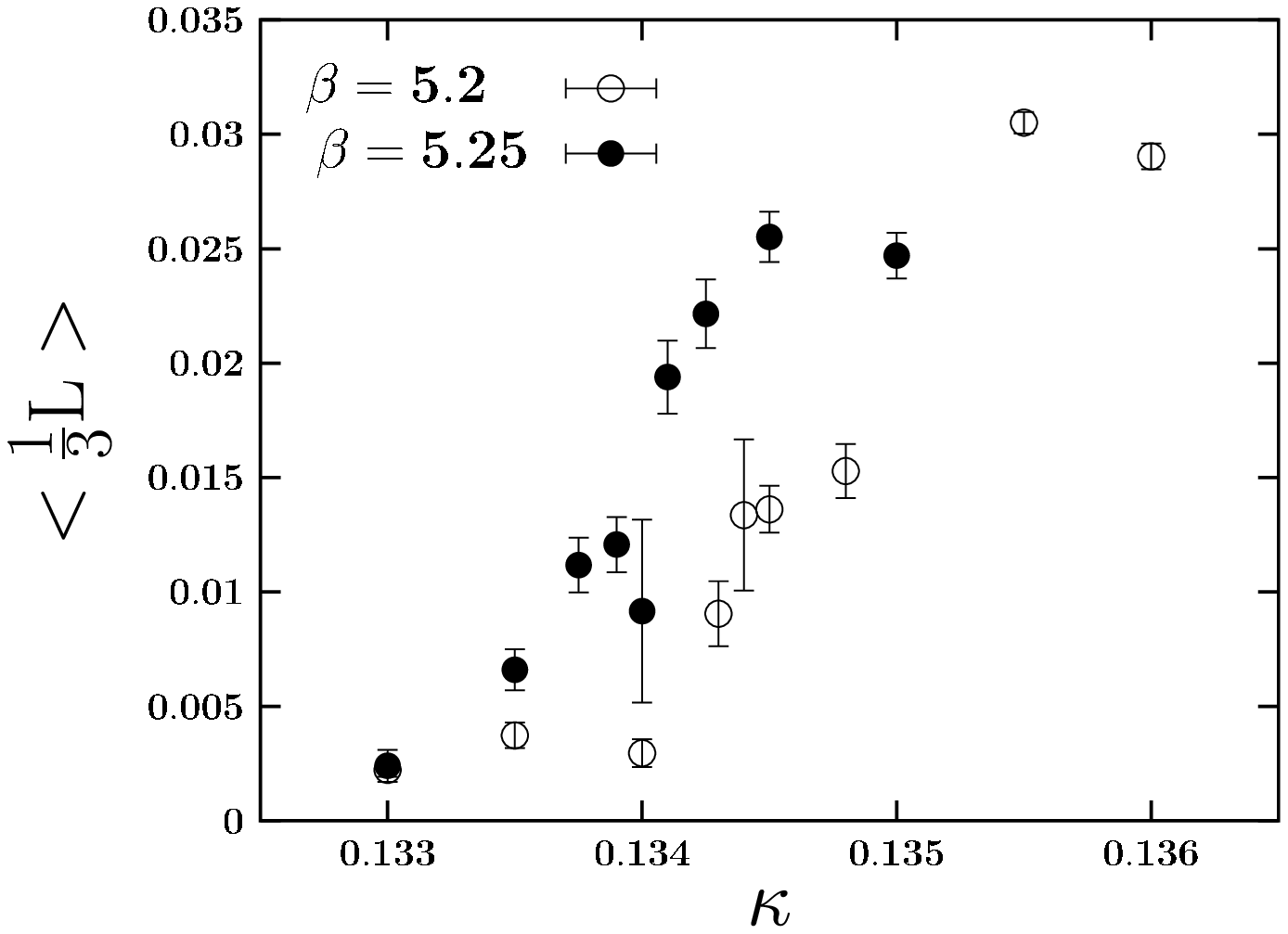}}
\vskip 2mm
\hbox{
\epsfxsize=6.cm
\hspace{0.2cm}\epsfbox{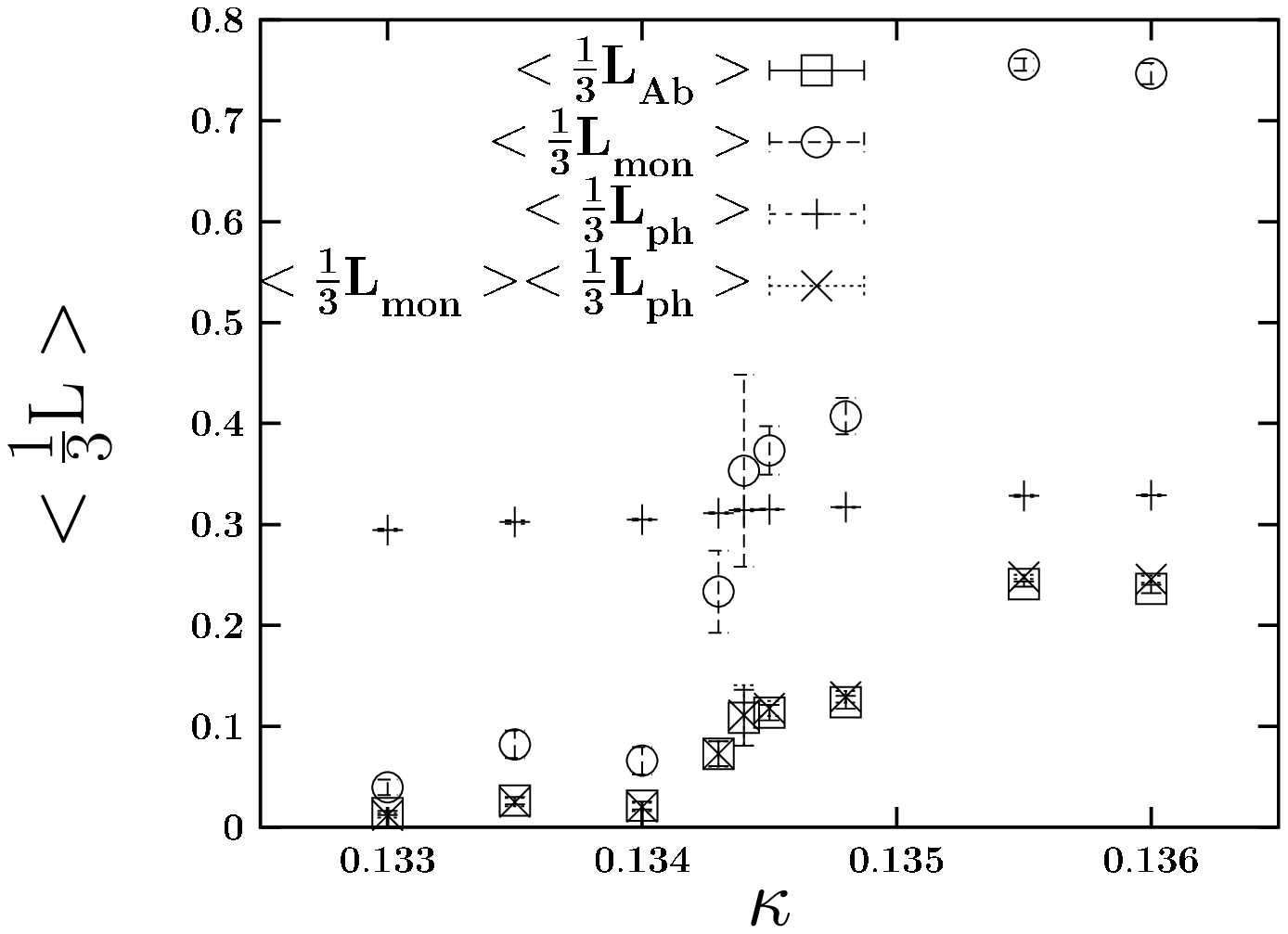}}
\vskip -8mm
\caption{Nonabelian Polyakov loop at both $\beta$'s (top), and
the abelian, monopole and photon Polyakov loops at $\beta=5.2$ (bottom).
\vskip -2mm
}
\label{ploop}
\vskip -5mm
\end{figure}
We measured the Polyakov loop and plaquette on every trajectory.
They were used to measure corresponding susceptibilities.
Depending on $\kappa$ every 5th, every 10th or every 20th trajectory
were used to compute Polyakov loops correlator
$\langle L_{\vec{x}} L^{\dagger}_{\vec{y}}\rangle $, where
$L_{\vec{x}} = \mbox{Tr}P_{\vec{x}}$,
$P_{\vec{x}}=\prod_{x_0} U_{0;x_0,\vec{x}}$.
The MA gauge was also fixed on these configurations. Simulated annealing
algorithm has been used to fix the gauge \cite{Bali:1994jg}. Abelian and monopole
Polyakov loops, their susceptibilities and correlators were measured on
gauge fixed configurations.
It turns out that for abelian and monopole observables
the signal/noise ratio was better than that for gauge invariant
nonabelian observables. This observation is in agreement with
results from quenched QCD at $T>0$ and both quenched and unquenched QCD at $T=0$.
For this reason we use abelian and monopole observables to determine
transition temperature and to study string breaking.
To obtain more precise results with our limited statistics we also employed
hypercubic blocking introduced recently~\cite{Hasenfratz:2001hp}.
\section{Transition temperature}
\begin{figure}[thb]
\hbox{
\epsfxsize=6.cm
\hspace{0.cm} \epsfbox{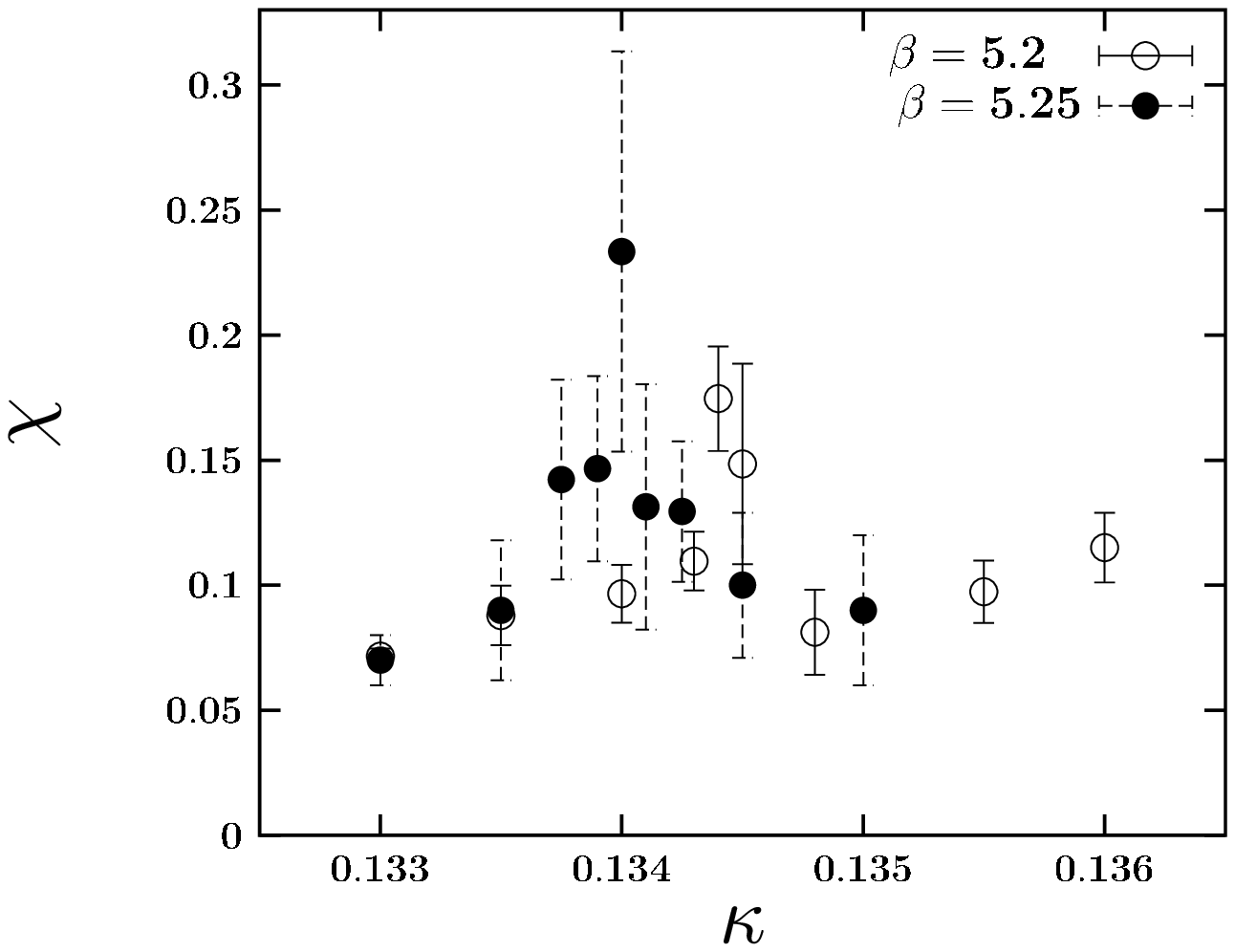}}
\vskip 2mm
\hbox{
\epsfxsize=6.cm
\hspace{0.cm} \epsfbox{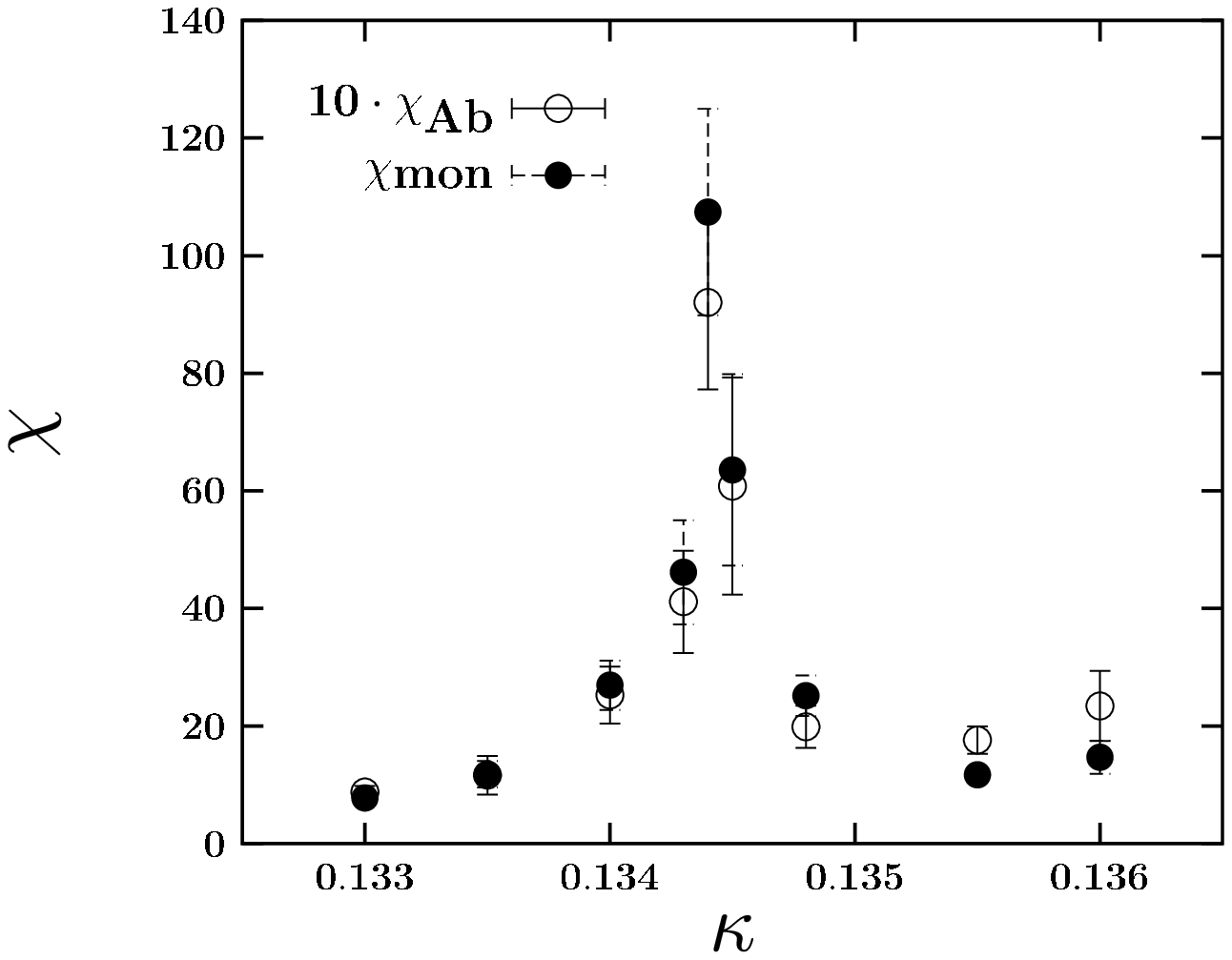}}
\vskip -6mm
\caption{Nonabelian Polyakov loop susceptibilities (top),
and abelian and monopole Polyakov loop susceptibilities at $\beta=5.2$
(bottom).
\vskip -4mm}
\label{nabsusc}
\vskip -4mm
\end{figure}
In Fig.\ref{ploop} we show results for average of various kinds of Polyakov
loops. One can see that $\langle L\rangle $ is a smooth function of $\kappa$.
In the bottom
part of this figure we show that as in quenched theory $\langle L_{Ab}\rangle  \approx
\langle L_{\mathrm{mon}}\rangle \langle L_{\mathrm{ph}}\rangle $ and
that $\langle L_{ph}\rangle $ does not show any changes at the transition.
We determined $\kappa_t$, the critical $\kappa$, from the maximum of the
Polyakov loop
susceptibility. We found  $\kappa_t=0.1344(1)$ at $\beta=5.2$
and $\kappa_t=0.1340(1)$ at $\beta=5.25$, see Fig.~\ref{nabsusc}.

An interpolation formula for $ r_0/a $ \cite{Booth:2001qp} gives
us $T_c r_0 = 0.54(2)$ and $0.56(2)$. Or, using $r^{-1}_0=394$
MeV, we obtain the critical temperature in physical units,
$T_c=213(10)$ and $220(5)$MeV. Using again the data from
\cite{Booth:2001qp} we get $m_{\pi}/m_{\rho}=0.78, 0.82$ at
the transition points. The susceptibilities for Abelian and
monopole Polyakov loops have maxima at the same $\kappa$, see
Fig.~2.

\begin{figure}[thb]
\hbox{
\epsfxsize=6.cm
\hspace{0.cm} \epsfbox{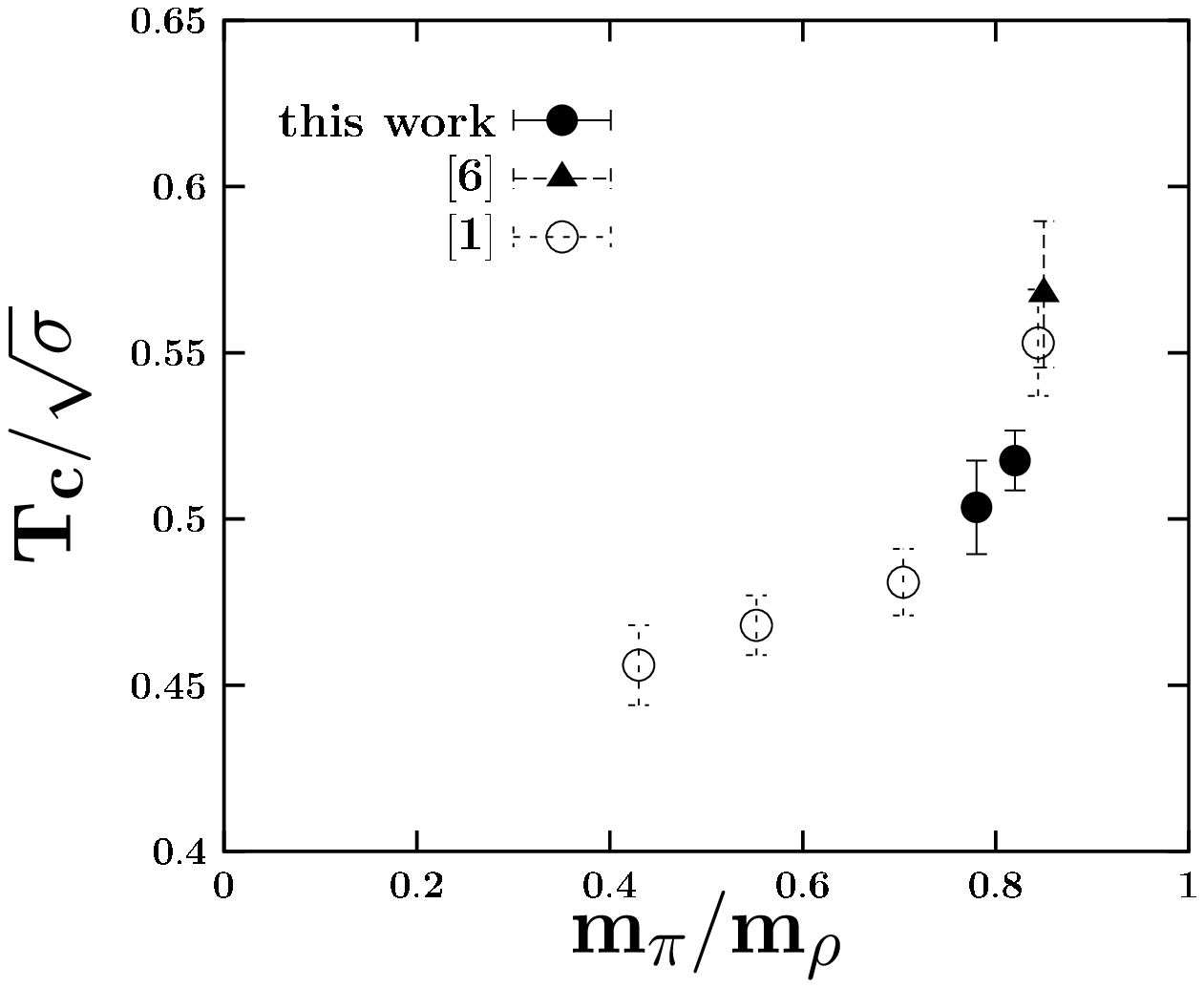}
}
\vskip -8mm
\caption{Transition temperature.
\vskip -10mm}
\label{tt}
\end{figure}

In Fig. \ref{tt} our results for transition temperature are shown in
comparison with those of Refs.~\cite{Karsch:2000kv,Edwards:1999mm}.
Thus our results are in a good qualitative agreement with the results
of the Bielefeld group.
\vskip -4mm

\vspace{5mm}
\section{String breaking}
\subsection{$T=0$}
At $T=0$ string breaking has been observed so far with mixed operators
including explicitly static-light meson state and no sign of string breaking
from Wilson loop has been found \cite{fk}.
The last fact is believed to be due to very poor overlap of the broken
string state with the Wilson loop operator.
Keeping two terms in the Wilson loop spectral representation, corresponding
to the string and two-meson states, one gets
\begin{equation}
  W(r,t) = C_V(r)e^{- (V_0+V_{str}(r) ) t} +  C_E(r)e^{-2E\cdot t},
\label{wloop}
\end{equation}
where $V_{str}(r)$ is the usual confining potential and $E$ is the static-light
meson energy. The overlap $C_V(r)$ is of $O(1)$ while $ C_E(r)$ might be small.
A quantitative estimate of this overlap has been suggested in
\cite{Aoki:1999sb}
and then derived in abelian projection approach~\cite{suzuki}:
$C_E(r) \sim e^{-2E\cdot r}$. We believe that
this is correct estimate at least in strong coupling, small $\kappa$
limit. If we assume that the real overlap is indeed of this order
then the effect of the second term in eq.~(\ref{wloop}) would become
detectable for both $r$ and $t$ of order of 2 fm, i.e. for very large
Wilson loops with very small value.
The real string breaking distance $r_{sb}$ is to be estimated from the equality
of two exponents in eq.~(\ref{wloop}):
$
 2m = \sigma \cdot r_{sb}- \pi \slash (12r_{sb}),
$
where $m=E - V_0/2 $ may be called the string breaking energy or
the effective quark mass \cite{Digal:2001iu}.
The Wuppertal group
result is $r_{sb}=2.3 r_0$ at $m_\pi/m_\rho=0.7$ \cite{Bali:2000vr}
and the CP-PACS result is $r_{sb}=2.2 r_0$  at $m_\pi/m_\rho=0.6$
\cite{Aoki:1999sb}.
Taking the string tension in full QCD from Ref.~\cite{Bali:2000vr},
$\sqrt{\sigma} r_0 = 1.15(1)$, we get
the effective quark mass $2m \sim 2.9/r_0 \sim 1.1 \mbox{GeV}$.
This is in a surprisingly good agreement with estimate
made in \cite{Digal:2001iu} using a different approach.

\subsection{$T>0$}

At non-zero temperatures the string breaking
has been observed in \cite{DeTar:1999qa}.
The heavy quark potential $V$ is related to the Polyakov loop
correlator,
$
e^{-V(r,T)/T}=\!\!
\langle L_{\vec{x}} L^{\dagger}_{\vec{y}}\rangle
\slash 9
$.
In the limit
$|\vec{x}-\vec{y}| \rightarrow \infty$,
$\langle L_{\vec{x}} L^{\dagger}_{\vec{y}}\rangle$ approaches the
cluster value $|\langle L\rangle|^2$,
where $|\langle L\rangle|^2 \neq 0$ because the global $Z_3$ symmetry
is broken by the fermions.

\begin{figure}[bth]
\vskip -8mm
\hbox{
\epsfxsize=6.cm
\hspace{-0cm} \epsfbox{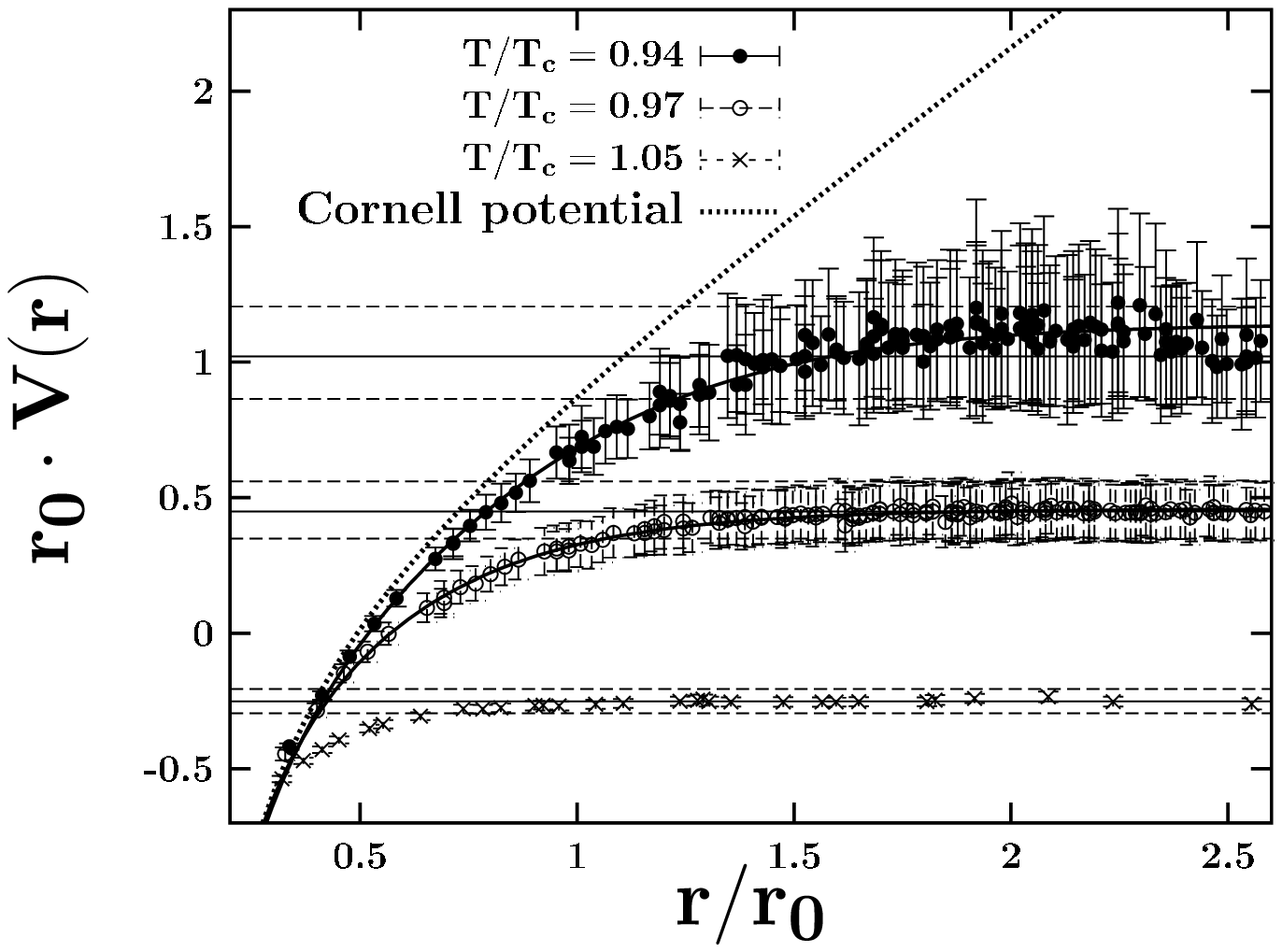}}
\hbox{
\epsfxsize=6.cm
\hspace{-0cm} \epsfbox{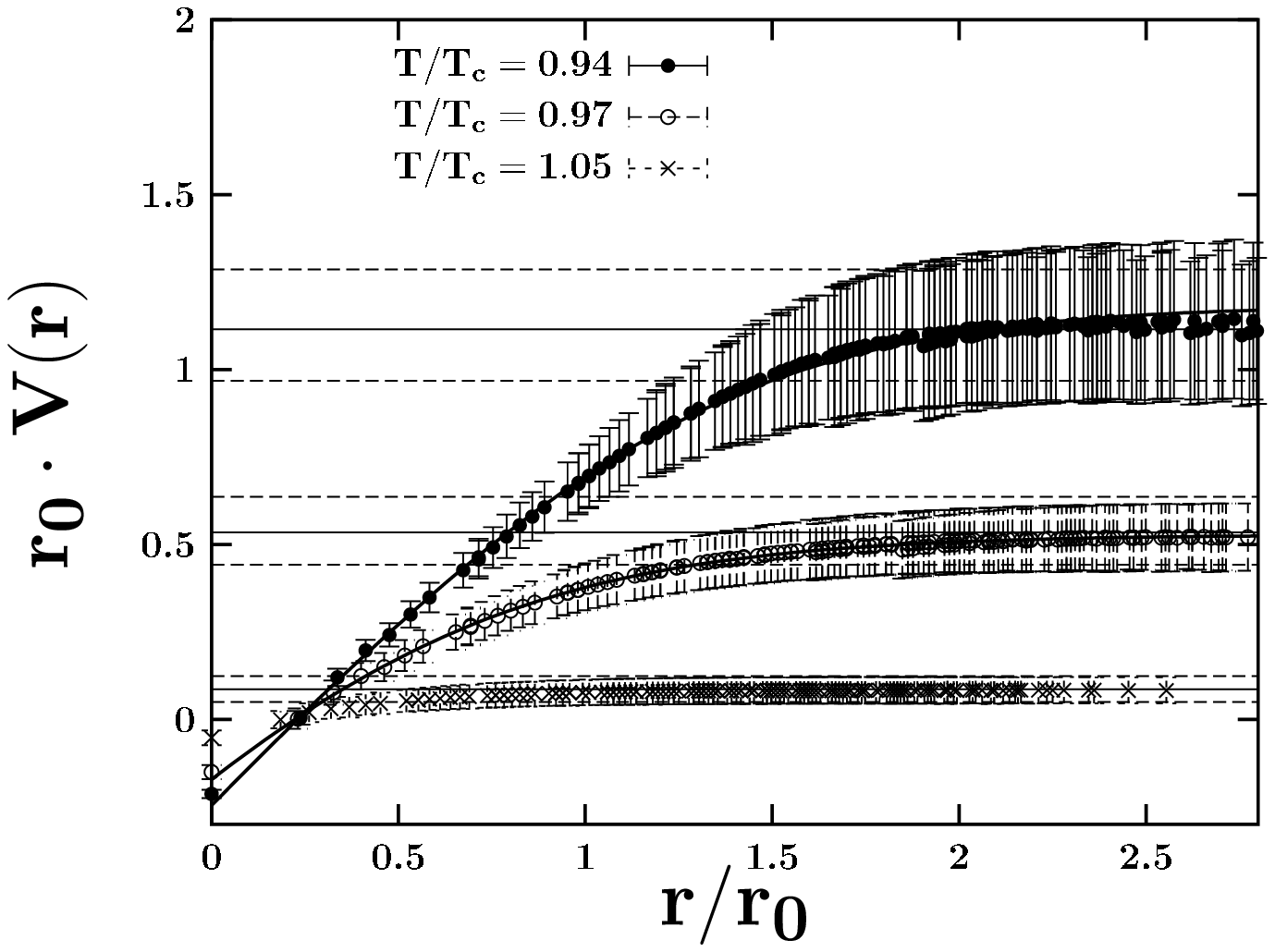}}
\vskip -6mm
\caption{Heavy quark potential from nonabelian (top) and
monopole (bottom) Polyakov loops, fitted by
eq.(\ref{two_exp}) and eq.(\ref{mon_fit}), respectively.
Horizontal lines show $-2r_0T\, \mbox{log}\langle L \slash 3\rangle$ with
the error bars.}
\label{nab_pot}
\vskip -8mm
\end{figure}
The spectral representation for the Polyakov loop correlator is
\cite{Luscher:2002qv}
$$
\langle L_{\vec{x}} L^{\dagger}_{\vec{y}}\rangle  = \sum_{n=0}^{\infty}
w_n e^{-E_n(r)/T},
$$
with integer $w_n$.
At $T=0$ we get $V(r,T)=E_0(r)$. In contrast, $V(r,T)$ at $T>0$
gets contributions from all
possible
states. In pure gauge theory
these are singlet and octet states and their excitations.
Assuming that the excitations have much higher energies
one can write
$$
e^{-V(r,T)/T} = \frac{1}{9} e^{-V_{sing}(r,T)/T} + \frac{8}{9}
e^{-V_{oct}(r,T)/T}\,.
$$
In full QCD we must take into account the excited states. As it was discussed
above  in full QCD at $T=0$ the singlet potential can be described by
string model potential up to some distance $r_{br}$.  Beyond this distance
another state becomes the ground state of the system  of two static quarks
which one can call broken string state or two heavy-light mesons state.
So there are two distinct states in the spectrum, one of which becomes
the ground state at proper distance. The situation must be similar
at small temperatures. We now assume that at temperatures
$T<T_c$ the Polyakov loop correlator can be described with the help
of these two states, namely string state and broken string (two meson)
state. In this paper we do not single out the singlet potential.
We understand that the contribution from the octet potential may
contaminate our results. The work on calculation of the singlet and
octet potential separately is in progress.
\begin{figure*}[tbh]
\hbox{
\epsfxsize=5.35cm
\hspace{-0.2cm} \epsfbox{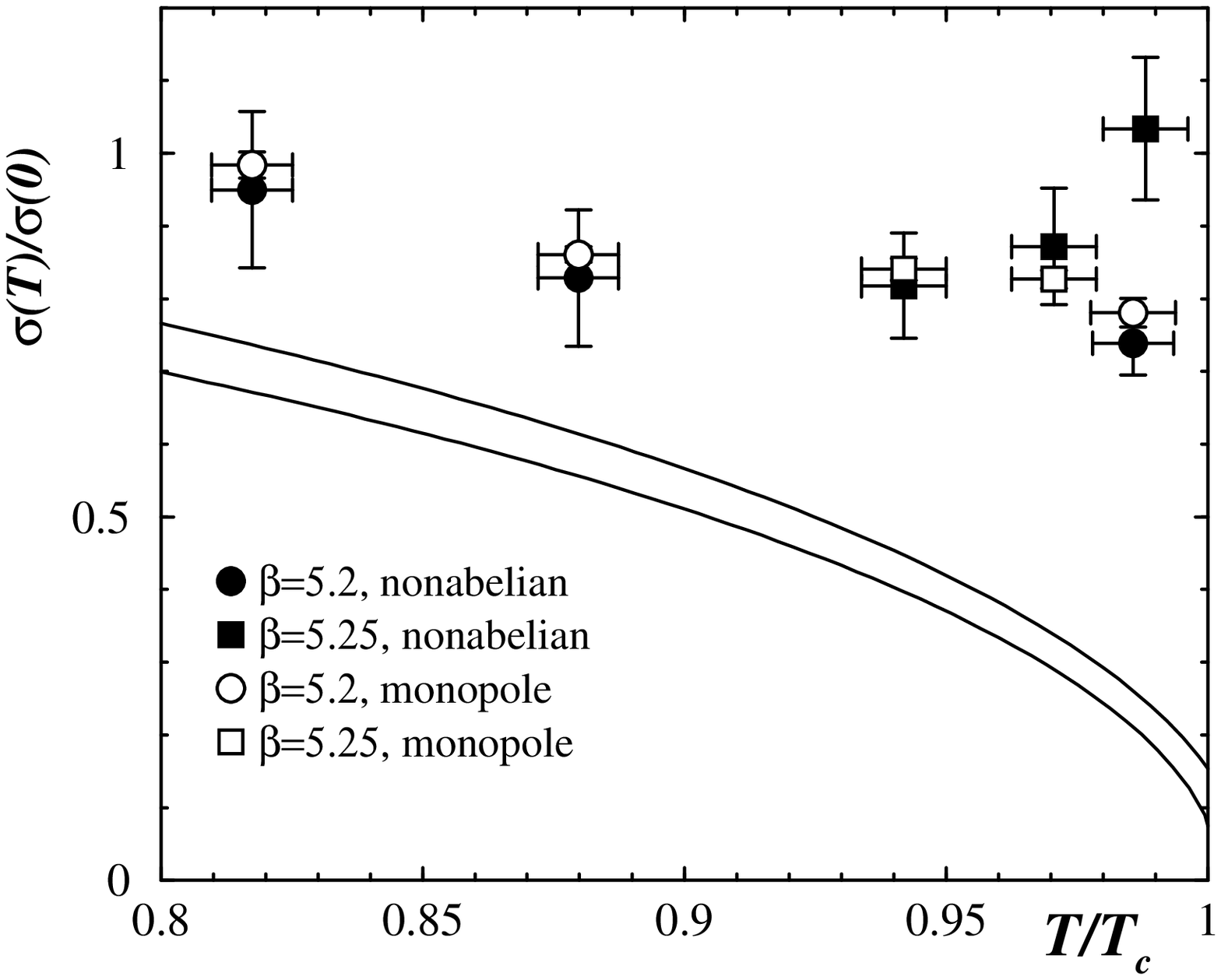}
\epsfxsize=5.35cm \hspace{-0.2cm} \epsfbox{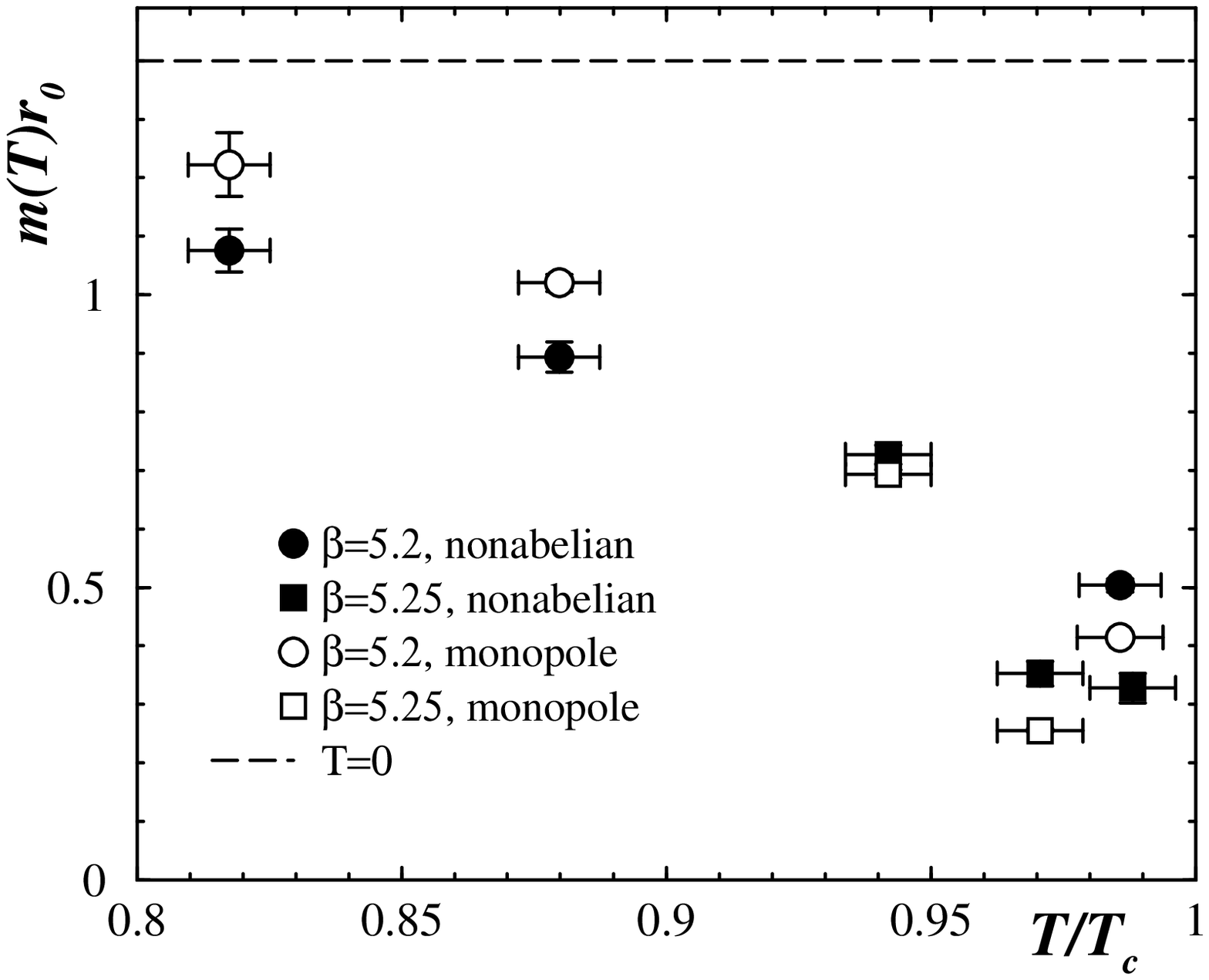}
\epsfxsize=5.35cm \hspace{-0.2cm} \epsfbox{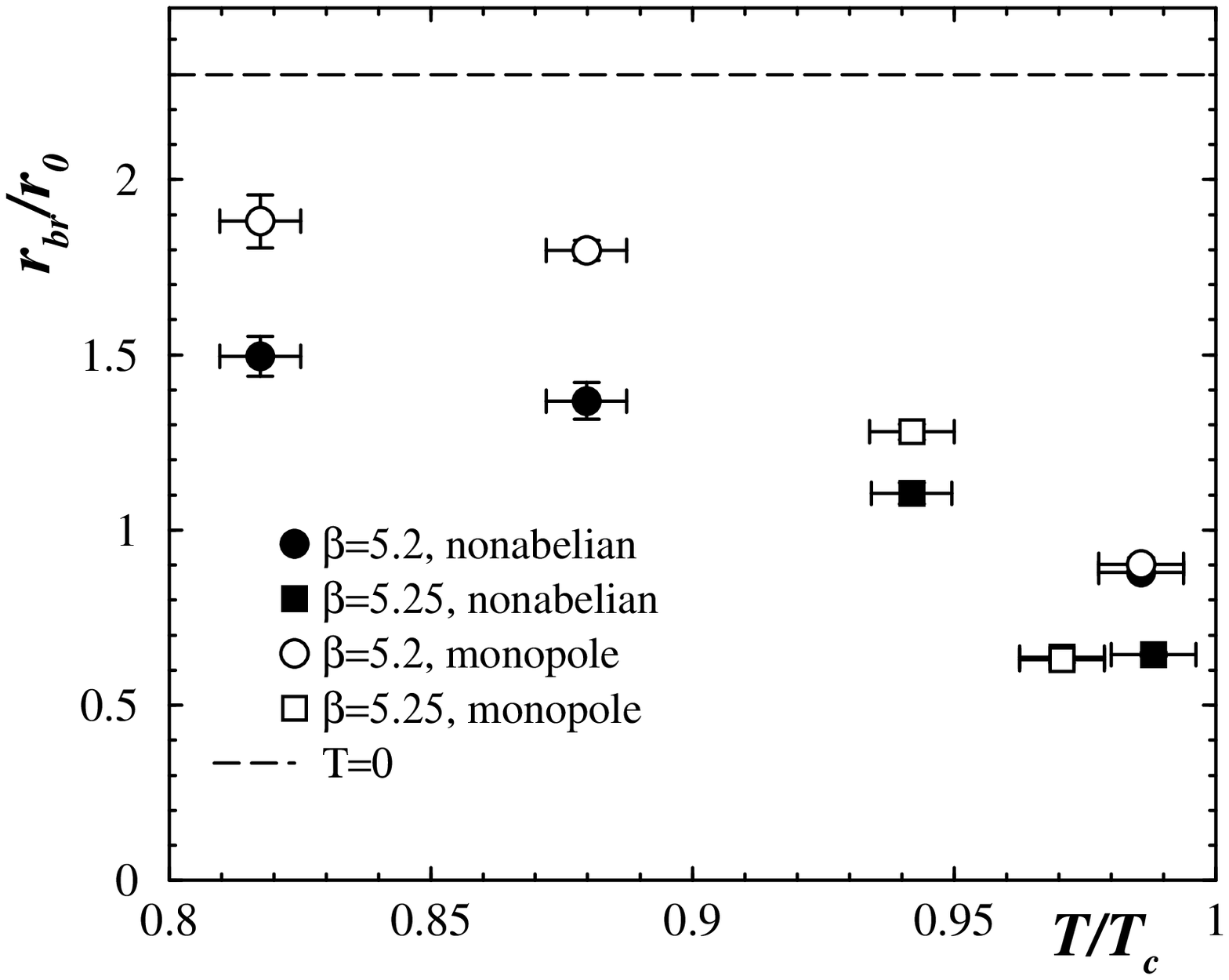}}
\vskip -8mm
\caption{Best fit parameters for nonabelian and monopole potentials
as functions of temperature. Solid lines on left figure show quenched
results~\cite{Kaczmarek:2000mm}.}
\label{sigma}
\vskip -2mm
\end{figure*}
We now adopt the following representation for the Polyakov loop
correlator
\vspace{-1mm}
\begin{eqnarray}
& & \hskip -6mm \frac{1}{9}\langle L_{\vec{x}} L^{\dagger}_{\vec{y}} \rangle =
 e^{-(V_0+V_{str}(r,T))/T} + e^{-2E(T) /T}\,,
\vspace{-1mm}
\label{two_exp}
\\
& & \hskip -6mm
V_{str}(r,T) = -\left(\frac{\pi}{12}-{1\over 6} {\rm arctan}(2 r T) \right)\,{1\over r}
+
\vspace{-1mm}
\label{pot_string}
\\
&&
\hskip -6mm
\Bigr(\sigma(T)+{2 T^2\over 3} {\mathrm {arctan}}{1\over 2r T}\Bigr)r +
\vspace{-1mm}
{T\over 2} \ln \Bigl(1+4 r^2 T^2\Bigr)\,,
\vspace{-1mm}
\nonumber
\end{eqnarray}
\vspace{-5mm}
\begin{equation}
\vspace{-1mm}
 E(T)=V_0/2 + m(T)\,,
\label{meff}
\end{equation}
\vspace{-1mm}
where $m(T)$ is effective quark mass at finite temperature. The
$T\neq0$ string potential (\ref{pot_string}) was calculated
in Ref.~\cite{Gao:kg}.
An alternative way to fit the static potential in full QCD at
finite temperature was proposed long  ago \cite{Karsch:1987pv}:
\beq
V(r,T)=\frac{\tilde{\sigma}}{\mu}(1-e^{-\mu r}) - \frac{\alpha}{r} e^{-\mu r}
\label{kms}
\eeq
with parameters $\tilde{\sigma}$ and $\mu$.
In (\ref{kms}) only $\mu$ \cite{Karsch:1987pv} or only
$\tilde \sigma$ \cite{Wong:2001uu} are temperature dependent.
We used function (\ref{kms}) to fit our data and compare with our model
eq. (\ref{two_exp})-(\ref{meff}).

Fig. \ref{nab_pot} shows the nonabelian static potential for some values of $T/T_c$.
The fit curves are two-state fits introduced above. We find good
agreement of the fit with lattice data.
As it has been already mentioned above to reduce statistical errors
we applied hypercubic blocking (HCB) \cite{Hasenfratz:2001hp}. We used
the same parameters of HCB procedure as in
\cite{Hasenfratz:2001hp}.
HCB helped to reduce the statistical errors by about factor 2 and
to improve the rotational invariance. The potentials without and
after HCB coincide within error bars up to a constant at all
distances $r$ apart from $r=a$ and $a\sqrt{2}$.

Parameters of our fit (\ref{two_exp}) are presented in Fig.~\ref{sigma}.
Our values for the ratio $\sigma(T)/\sigma(0)$
are higher than those obtained in quenched QCD in
Ref.~\cite{Kaczmarek:2000mm}.
Our values for the quark effective
mass are also higher than those obtained in \cite{Digal:2001iu}.
Having parameters of the potential
determined we can calculate the string breaking distance $r_{sb}$
using the relation
$V_{str}(r_{sb},T) = 2m(T)$.
{}From Fig. \ref{sigma} one can see that $r_{sb}$ decreases down to values
$\sim 0.3$ fm  when temperature approaches critical value. Our fit using
$V_{str}$ is probably not valid when $r_{sb}$ becomes so small. It still
can be valid for $T/T_c < 0.95$ when $r_{sb} > 0.5 fm$.

The quality of our data for $\langle L_{\vec{x}}
L^{\dagger}_{\vec{y}} \rangle$ does not allow us to distinguish
between the fits eq.~(\ref{two_exp}-\ref{meff}) and (\ref{kms}).
However, the effective string tension in the fit (\ref{kms}) is
unphysically large, $\tilde \sigma r_0^2 \sim 6$.

We calculated also the potential after abelian projection. It is known
from quenched QCD studies that the monopole part of the abelian
gauge field gives a potential which is linear down to very small
distances, $i.e.$ the Coulomb coefficient is compatible with zero.
We observed similar behaviour at small distances as is seen
from Fig. \ref{nab_pot}.
At large distances the monopole potential shows screening as
the nonabelian potential does. We fit the potential
similarly to eq.(\ref{two_exp}):
\begin{equation}
\!\!{\frac{1}{9}\langle L_{\vec{x}}
L^{\dagger}_{\vec{y}}\rangle}^{\mathrm{mon}}\!\!\!\!\!\!\!\!
= e^{-(V_0^{\mathrm{mon}}+V_{str}^{\mathrm{mon}}(r,T))/T}
\!\! + e^{-2E^{\mathrm{mon}}(T)/T}
\label{mon_fit}
\end{equation}
$V_{str}^{\mathrm{mon}}
= \sigma^{\mathrm{mon}} \, r$ and
$E^{\mathrm{mon}}(T)=V_0^{\mathrm{mon}}+m^{\mathrm{mon}}(T)$. The fit is
shown in Fig.\ref{nab_pot} by the solid line.

\vskip -4mm
\begin{figure}[thb]
\hbox{
\epsfysize=4.5cm
\epsfxsize=5.5cm
\hspace{0.cm} \epsfbox{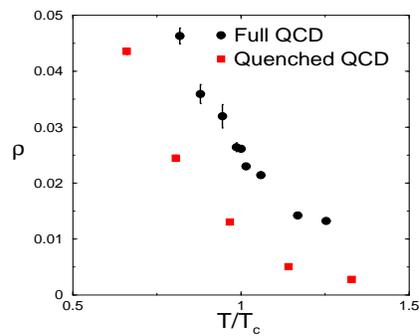}}
\vskip -10mm
\caption{Density of magnetic currents.}
\label{dens}
\vskip -6mm
\end{figure}
\section{Monopole Dynamics}
We studied monopole density $\rho$ and asymmetry in the monopole
density $\eta = (\rho_t-\rho_s) \slash (\rho_t+\rho_s)$
where $\rho_t$($\rho_s$) is density of time-like (space-like)
magnetic currents.  In Fig. \ref{dens} we compare the monopole density in
full and quenched QCD. Our results for quenched QCD were obtained on
$16^3 \cdot 8$ lattice.
The density in full QCD is substantially higher than that in quenched
theory in agreement with our earlier results at $T=0$ \cite{Bornyakov:2001nd}.
The behaviour of $\eta$ is qualitatively similar in both theories,
i.e. $\eta$ is
close to
zero below transition and increases with temperature above $T_c$.
\section{Conclusions and Outlook}
Our results for $T_c$ obtained on $16^3 \cdot 8$ lattice at
$\frac{m_{\pi}}{m_{\rho}} \sim 0.8$ and lattice spacing small in
comparison with  work of Bielefeld group are in agreement with
their results~\cite{Karsch:2000kv}. To test finite size effects
and to determine transition temperature closer to the chiral limit
we are planning to make simulations on $24^3\cdot 8$ and
$24^3\cdot 10$ lattices.

Heavy quark potential has been measured in both phases. We
introduced two-state parameterization
eq.(\ref{two_exp}-\ref{meff}) for the heavy quark potential in
confinement phase and found good agreement with numerical data
obtained in that phase. Using this parameterization we computed
string tension, quark effective mass and string breaking distance.
We found that the ratio $\sigma(T)/\sigma(0)$ decreases toward the
value 0.7 when $T$ approaches $T_c$. This value is substantially
higher than the value obtained in quenched QCD. Whether this is
due to systematic effects (meson-meson interaction, contribution
of the octet potential, and some others) or is a physical effect
deserves further study. Our results for the quark effective mass
show good qualitative agreement with earlier
results~\cite{Digal:2001iu}. The results for string breaking
distance imply that our fit might be inappropriate for
temperatures $T/T_c>0.95$.

We observed string breaking at $T < T_c$  also from abelian and monopole
Polyakov loop correlators. We found that string tension, quark effective mass
and string breaking distance calculated from the monopole Polyakov
loops correlators are in agreement with the values obtained from
the nonabelian correlators. Thus we found abelian and monopole
dominance in the confinement phase of full QCD.

\section{Acknowledgements}
We are very obliged to  the staff of the Joint Supercomputer
Center at Moscow and especially to A.V. Zabrodin for the help in
computations on supercomputer MVS 1000M. This work is partially
supported by grants INTAS-00-00111, RFBR 02-02-17308, RFBR
01-02-117456, RFBR 00-15-96-786 and CRDF award RPI-2364-MO-02.
V.B. and M.Ch. (grant No. P01023) are supported by JSPS Fellowships.


\begin{thebibliography}{99}
\bibitem{fk}
New developments are
reported by Ph.de For\-crand and S. Kratochvila,
these Proceedings.
\end{thebibliography}
\end{document}